\documentclass[final,5p,times,twocolumn]{elsarticle}
\usepackage{amsmath}
\usepackage{amssymb}
\usepackage{amsfonts}

\usepackage{graphicx}
\graphicspath{{figs/}}

\usepackage{booktabs}
\usepackage{multirow}
\usepackage{array}
\usepackage{xcolor}
\usepackage{algorithm}
\usepackage{algorithmic}
\usepackage{url}
\usepackage{tabularx}
\usepackage{ragged2e}
\usepackage{makecell}

\usepackage[hidelinks]{hyperref}

\usepackage{lineno}

\journal{Advanced Engineering Informatics}

\begin{document}

\begin{frontmatter}


\title{LLMs for Agentic Home Energy Management}


\author[aff1]{Sokipriala Jonah\corref{cor1}}
\ead{jonahs@uni.coventry.ac.uk}

\author[aff2]{Queen Moses}
\author[aff1]{Abiola Babatunde}

\author[aff1]{Michael Ajao-Olarinoye}
\author[aff3]{Daniel Bammeke}

\cortext[cor1]{Corresponding author}


\affiliation[aff1]{
    organization={Centre for Computational Science and Mathematical Modelling,
    Coventry University},
    city={Coventry},
    country={United Kingdom}
}

\affiliation[aff2]{
    organization={Heriot-Watt University},
    city={Edinburgh},
    country={United Kingdom}
}

\affiliation[aff3]{
    organization={Aston University},
    city={Birmingham},
    country={United Kingdom}
}

\begin{abstract}
Home Energy Management Systems (HEMS) can reduce residential electricity
costs, but many require users to express everyday preferences as technical
constraints. This paper presents a tool-calling ReAct agent that converts
natural-language requests into schedules for multiple household appliances
using half-hourly Octopus Agile prices, weather forecasts, photovoltaic
generation estimates, and household demand data. Five LLM backends are
evaluated against a mixed-integer linear programming benchmark across
dynamic tariff conditions, constraint conflicts, weather-aware scheduling,
and a seven-day rolling deployment.
Native function calling achieves high scheduling success and near-optimal
cost on ordinary tariff days, whereas text-parsed actions reduce reliability.
Constraint-conflict testing shows that low cost does not guarantee safe or
feasible behaviour. Claude Sonnet~4.6 performs best in power-cap and
infeasibility scenarios, while Qwen-3 achieves higher overall constraint
compliance than GPT-4o-mini. The evaluation also identifies fabricated
schedules, failed commitments, and reasoning-to-action failures in which
models explain a deadline correctly but commit an invalid schedule.
Weather-aware scheduling reduces cost and increases solar self-consumption
under overcast conditions, but provides limited or adverse economic value
under some dynamic-price regimes. Across the evaluated seven-day period,
the agents capture 96.7--98.0\% of the savings available between an
off-peak timer and the MILP oracle and outperform the rule-based policies.
The results support LLM-based HEMS orchestration provided that every
committed schedule is checked by an independent deterministic feasibility
validator before actuation.Code and a live demonstration are available at \url{https://github.com/sokistar24/ecohome-experiments} and \url{https://www.ecohomeagent.com/}.

\end{abstract}


\end{frontmatter}


\section{Introduction}
\label{sec:introduction}

The growing use of variable renewable sources, particularly wind and solar
power, introduces greater variability into electricity supply because their
output changes with weather conditions. As a result, electricity systems
increasingly depend on demand-side flexibility, whereby homes and businesses
shift some of their electricity use to periods when supply is more available
or less expensive. This flexibility can help maintain system reliability
while limiting energy costs
\cite{iea2023grids,cabot2024demand}.

The International Energy Agency estimates that global demand-response
capacity must increase roughly tenfold between 2020 and 2030, with buildings
and residential electric vehicles (EVs) expected to provide a substantial
share of this potential
\cite{iea2023demandresponse,duan2022characterizing,felez2025advanced}.
Many residential loads, including EV charging, washing machines,
dishwashers, heating, and cooling, can be shifted to different times of the
day without materially reducing the service delivered to occupants
\cite{golmohamadi2024demand,d2022exploiting}.

Home Energy Management Systems (HEMS) support this flexibility by
coordinating household loads in response to electricity prices, local
generation, and user requirements. However, despite their potential to
reduce costs and support demand response, adoption remains limited. One
important barrier is that conventional HEMS often require residents to
translate everyday preferences, such as when an EV must be ready or when an
appliance may operate, into technical scheduling parameters and constraints
\cite{shimoda2021evaluating,khafiso2024barriers,
michelon2025llm}.

Large language models (LLMs) could reduce this interaction burden by
allowing users to express such requirements in natural language. When
combined with external tools, an LLM can retrieve tariff and weather data,
interpret household constraints, invoke scheduling functions, commit
actions, and explain the resulting recommendation
\cite{xu2026theagentcompany,
yao2023react,aksitov2023rest}. Retrieval-augmented generation can further
ground general energy advice in a curated knowledge base. In this role, the
LLM does not merely answer questions; it acts as an orchestration layer
between the user, household data, external information sources, and
executable scheduling tools.

Recent work has begun to explore this role in residential and building
energy management. Existing systems have used LLMs to provide
conversational energy advice, extract scheduling parameters from user
requests, interface with mathematical optimizers or model-predictive
controllers, and coordinate appliance schedules
\cite{jung2026multi,elmakroum2026agentic,raghavan2026cost}. In most cases, however, the LLM remains an
interface or advisory component, while a conventional optimizer or
controller makes the final scheduling decision.

El~Makroum \emph{et al.} \cite{elmakroum2026agentic} move beyond this
interface-only role by allowing an LLM orchestrator to commit schedules for
a washing machine, dishwasher, and EV charger. Their hierarchical system
matched mixed-integer linear programming (MILP) optima with
Llama-3.3-70B. However, Qwen-3-32B and GPT-OSS-120B failed to coordinate
all three appliances despite achieving perfect performance on
single-appliance tasks. These findings demonstrate that autonomous
LLM-based scheduling is feasible, but they also leave several questions
unresolved:

\begin{enumerate}
  \item \textbf{Model and interface effects.} Existing autonomous results
  combine open-source models, hierarchical decomposition, and text-parsed
  actions, making it difficult to identify the source of coordination
  failures.

  \item \textbf{Price-only scheduling.} Wholesale price alone does not
  capture household net cost when rooftop photovoltaic (PV) generation and
  export revenue affect the preferred schedule.

  \item \textbf{Limited evaluation horizon.} Single-day testing does not
  establish robustness across price-volatility regimes or sustained use.

  \item \textbf{Constraint safety.} Prior evaluations do not test cases in
  which low-cost schedules conflict with deadlines, calendars, power caps,
  or physical feasibility.
\end{enumerate}

We address these gaps using a single tool-calling ReAct agent evaluated
with five commercial and open-source LLMs under a common set of tools,
prompts, and scenarios. The system uses half-hourly Octopus Agile retail
prices in Great Britain, weather forecasts, PV generation estimates,
household usage data, and a retrieval-augmented knowledge base. Native
function calling and text-parsed actions are compared directly, while an
extended mixed-integer linear program (MILP) provides the scheduling
ground truth.

\subsection{Contributions}
\label{sec:contributions}

This paper makes four contributions:

\begin{itemize}
  \item \textbf{C1: Multi-model, multi-regime benchmark.} We compare
  GPT-4o-mini, Gemini~2.5 Flash, Claude Sonnet~4.6, Llama-3.3~70B, and
  Qwen-3~32B across 12 Agile tariff days, two action interfaces, and more
  than 1,100 runs. The evaluation reports success, exact optimality, cost
  gaps, latency, token use, and cost per successful schedule, followed by
  a seven-day rolling deployment against MILP and rule-based baselines.

  \item \textbf{C2: Constraint-conflict stress testing.} We introduce
  scenarios covering deadline conflicts, household power caps, irregular
  calendars, infeasible requests, conflicting user instructions, and tool
  failures. The suite measures constraint violations, infeasibility
  reporting, recovery, compliance cost, and trace-level failure modes.

  \item \textbf{C3: Weather-aware PV co-optimization.} We extend the
  objective from price-only scheduling to net cost with PV
  self-consumption and export revenue. An extended joint MILP benchmarks
  the resulting schedules, including sensitivity to PV forecast error.

  \item \textbf{C4: Reasoning-to-action validation.} We identify cases in
  which a model explains a deadline correctly but commits a violating
  schedule. This motivates a model-independent feasibility validator that
  checks committed actions before device actuation.
\end{itemize}
\section{Related Work}
\label{sec:related}

\subsection{Operational Foundations of Home Energy Management}

Building decarbonization research has traditionally considered physical
interventions such as passive-house retrofits, improved envelopes, and
whole-building renovation. These measures can provide durable energy savings,
but they often require capital investment, construction work, and substantial
information exchange between households and technical specialists
\cite{welch2023passive,wise2025retrofit,ibrahim2024retrofitting}. Home Energy
Management Systems (HEMS) provide an operational complement by coordinating
flexible demand in response to electricity prices, local generation, grid
signals, and household requirements. Their value therefore depends not only on
finding low-cost periods, but also on representing appliance durations,
deadlines, comfort requirements, and connection limits correctly
\cite{golmohamadi2024demand,khafiso2024barriers}.

Conventional HEMS use rule-based control, mathematical programming, model
predictive control (MPC), or reinforcement learning (RL). Rule-based approaches
are transparent and straightforward to deploy, and carefully engineered rules
can improve energy, comfort, and indoor-air-quality performance. However, their
fixed logic is difficult to adapt when future prices, weather, occupancy, or
appliance requests change \cite{faulkner2023guideline36}. MPC addresses this
limitation by repeatedly optimizing over a forecast horizon while enforcing
explicit physical and comfort constraints. It is therefore well suited to
price-responsive building operation, but its development requires system
models, objective design, constraint formulation, solver integration, and
specialist tuning \cite{drgona2020mpc,michailidis2025model,zong2017challenges}. RL can learn adaptive control policies
without an explicit analytical model, yet practical deployment remains
constrained by data requirements, limited interpretability, and the need to
demonstrate safe behaviour outside the training distribution
\cite{nagy2023rl,zhang2022building,yu2021review,li2026mcp}.

These methods are effective once objectives and constraints have been encoded,
but that encoding remains a barrier for household users. A resident may state
that an EV must be ready before work, that wet appliances should avoid a quiet
period, or that solar generation should be used where economical. Conventional
optimizers do not directly translate such requests into executable parameters.
This creates a distinct role for LLMs as natural-language reasoning and
orchestration layers. It also creates a safety boundary: flexible interaction
may be delegated to an LLM, while feasibility and physical constraints should
remain independently checkable.

\subsection{LLMs and Agentic Workflows in Energy and Buildings}

Agentic AI systems extend conversational LLMs by allowing the model to
decompose objectives, select tools, observe external results, and coordinate
multi-step actions \cite{sapkota2026agentic,hosseini2025agentic}. ReAct is a
widely adopted implementation pattern because it interleaves reasoning with
tool invocation and observation, producing traces that can be inspected during
error analysis \cite{yao2023react}. Native function calling further constrains
actions to structured tool names and arguments, reducing dependence on parsing
free-form responses. This distinction is important in energy applications,
where a fluent recommendation is not equivalent to an executable control
action.

LLMs have been studied in the energy domain as forecasting assistants,
optimization and code-generation tools, diagnostic systems, and decision
support models. Reviews identify grounding, domain adaptation, and external
tool use as central requirements because energy decisions depend on numerical,
temporal, and physical context \cite{majumder2024exploring}. For demand-side
management, LLMs have been used to formulate or support EV and flexible-load
problems \cite{zhang2025dsm,niu2025ev,fan2026spatiotemporal}. In building operation, customized models have
combined task routing with domain knowledge for fault diagnosis
\cite{chen2025oam,mirshekali2025review}, while other work has assessed LLM recommendations for
residential retrofit decisions \cite{shu2025retrofit}. Agentic HVAC control has
also been evaluated in occupied offices, demonstrating that LLM-mediated
control can improve energy and comfort outcomes in an operational setting
\cite{sawada2025office}. These applications establish broad capability, but
most address a bounded analytical task or a single building subsystem rather
than coordinated residential appliance scheduling.

A parallel research direction uses LLMs to automate engineering workflows.
EPlus-LLM translates natural-language building descriptions into EnergyPlus
models, reducing the manual effort required to construct simulation inputs
\cite{jiang2024eplusllm}. Ren \emph{et al.} extend this idea to controller
development in AgentiControl-MPC, where six specialized agents perform data
interpretation, predictive-model design, MPC formulation, coding, execution,
and correction \cite{ren2026agenticontrol}. In repeated runs, their multi-agent
system achieved a 93.3\% controller-generation success rate, compared with
66.7\% for a single-agent baseline; the selected controller also reduced energy
cost and thermal discomfort relative to the baseline controller. This is an
important demonstration of LLMs as engineering automation tools. However, the
LLM agents generate and debug a conventional MPC controller, whereas the MPC
remains responsible for runtime control. The present work instead evaluates an
LLM that directly selects and commits appliance schedules, making action
validity and constraint compliance part of the agent evaluation itself.

\subsection{LLM-Enabled Residential Energy Management}

Within smart homes, LLMs have been used to generate automation rules from
natural language \cite{giudici2025homeassistant}, combine household and IoT
context for adaptive energy decisions \cite{li2025smart}, and provide
conversational energy advice without direct actuation
\cite{gkalinikis2025rhea}. Other systems retain a conventional optimization
layer. Michelon \emph{et al.} extract scheduling parameters from dialogue and
pass them to a HEMS optimizer \cite{michelon2025llm}, while Raghavan and
Giridhar use an LLM as a natural-language interface to an MPC-based residential
battery scheduler \cite{raghavan2025mpc}. These architectures reduce the user
interaction burden while preserving deterministic optimization, but they do
not test whether the LLM itself can coordinate and commit several appliance
actions.

El~Makroum \emph{et al.} move beyond the interface-only role by allowing an LLM
orchestrator to commit schedules for a washing machine, dishwasher, and EV
charger \cite{elmakroum2026agentic}. Their hierarchical system matched the MILP
optimum with Llama-3.3-70B on Austrian day-ahead prices, while two other
open-source models failed to coordinate all appliances despite succeeding on
individual tasks. The study therefore establishes the feasibility of
autonomous LLM scheduling, but it also leaves model, interface, and
architecture effects entangled. Its evaluation uses one wholesale-price day,
does not incorporate weather, rooftop PV, household base load, or export
revenue, and does not examine infeasible requests or conflicts involving
deadlines, calendars, and household power limits.

This work addresses those gaps under a common single-agent architecture. Five
commercial and open-source models are compared using both native function
calling and text-parsed actions across multiple retail-tariff regimes. The
objective is extended from price-only scheduling to household net cost with
weather-informed PV generation and export revenue. The evaluation further
includes constraint-conflict stress tests, repeated runs, statistical analysis,
and a seven-day rolling benchmark against MILP and rule-based policies. The aim
is not to replace deterministic optimization in every HEMS component, but to
identify when an LLM can serve as a reliable orchestration layer and where
model-independent validation is required before actuation.

\begin{table*}
\caption{Positioning Relative to Closely Related LLM-Based Residential and Building Energy Systems}
\label{tab:positioning}
\centering
\footnotesize
\renewcommand{\arraystretch}{1.18}
\begin{tabularx}{\textwidth}{@{}
>{\RaggedRight\arraybackslash}p{2.35cm}
>{\RaggedRight\arraybackslash}X
>{\centering\arraybackslash}p{1.65cm}
>{\centering\arraybackslash}p{2.1cm}
>{\centering\arraybackslash}p{2.3cm}
>{\RaggedRight\arraybackslash}X
@{}}
\toprule
\textbf{Work} &
\textbf{LLM role / architecture} &
\textbf{Weather / PV} &
\textbf{Evaluation setting} &
\textbf{Constraint conflicts} &
\textbf{Objective / decision maker} \\
\midrule

Michelon \emph{et al.}~\cite{michelon2025llm} &
Dialogue-based parameter extraction for an external HEMS optimizer &
No &
Dialogue and optimizer evaluation &
No &
Optimizer commits the schedule \\

Raghavan \& Giridhar~\cite{raghavan2025mpc} &
Natural-language interface to an MPC battery scheduler &
No &
Residential battery case study &
No &
MPC commits control actions \\

Ren \emph{et al.}~\cite{ren2026agenticontrol} &
Six-agent workflow that generates and debugs an MPC controller &
Weather data; no PV scheduling objective &
15 multi-agent and 12 single-agent runs in BOPTEST &
MPC constraints; no LLM conflict stress suite &
Generated MPC controls heat-pump cost and comfort \\

Gkalinikis \emph{et al.}~\cite{gkalinikis2025rhea};
Li \emph{et al.}~\cite{li2025smart} &
Conversational advice or contextual smart-home decision support &
Limited / system dependent &
Advisor or prototype evaluation &
No dedicated suite &
Advice or high-level recommendations \\

El~Makroum \emph{et al.}~\cite{elmakroum2026agentic} &
Hierarchical LLM orchestrator with appliance-specialist agents &
No, prices only &
1 day, 75 runs &
Not tested &
LLM commits schedules against ENTSO-E wholesale prices \\

\textbf{This work} &
\textbf{Single tool-calling agent evaluated with five LLM backends} &
\textbf{Yes, weather-informed PV and household load} &
\textbf{12 tariff days, 1{,}100+ runs, and seven-day deployment} &
\textbf{Dedicated deadline, cap, calendar, infeasibility, instruction, and tool-failure suite} &
\textbf{LLM commits schedules; extended MILP provides ground truth} \\

\bottomrule
\end{tabularx}
\end{table*}

\section{System Architecture}
\label{sec:architecture}

\subsection{Agent Workflow}

\begin{figure*}
\centering
\includegraphics[width=\linewidth]{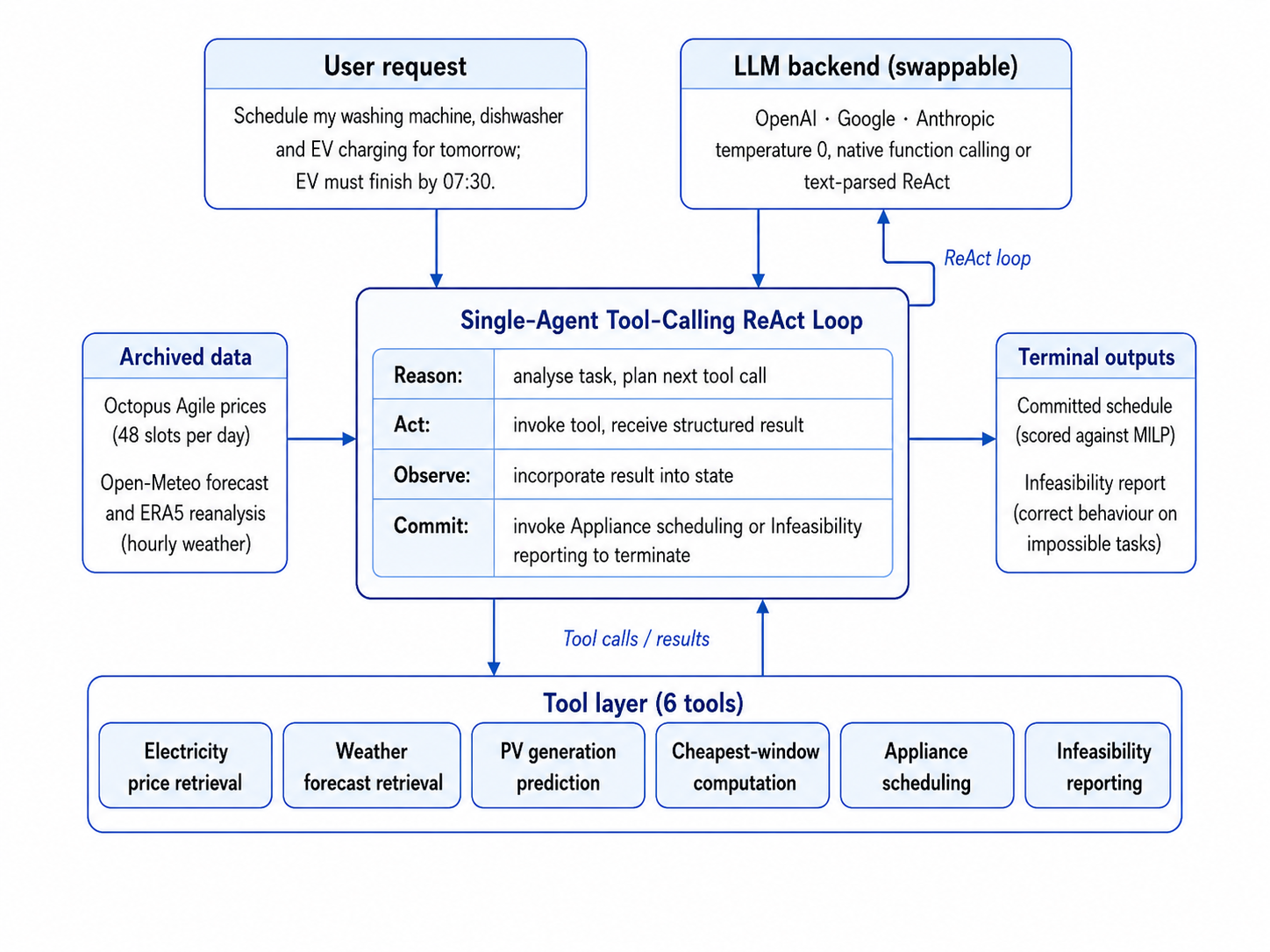}
\caption{System architecture. A single ReAct agent retrieves price and
weather data, invokes structured scheduling tools, and iterates until it
commits a schedule or reports infeasibility.}
\label{fig:architecture}
\end{figure*}

Figure~\ref{fig:architecture} shows the proposed architecture. A single
LangGraph ReAct agent receives a natural-language scheduling request and
iterates between tool invocation and observation until it commits a
schedule or reports infeasibility. It then returns the selected schedule,
estimated cost, and supporting explanation.

The system prompt defines the agent's role, required data sources,
constraint priorities, and response format. Unlike the hierarchical
orchestrator in \cite{elmakroum2026agentic}, the proposed system uses no
per-appliance specialist agents. This flat design allows the evaluation to
test whether structured tool use is sufficient for multi-appliance
coordination without manual task decomposition.

\subsection{Tool Layer}
\label{sec:tools}

The experimental agent has six tools:

\begin{itemize}
  \item \textbf{Electricity price retrieval} returns 48 half-hourly
  Octopus Agile import prices for a specified date
  \cite{octopusapi}.

  \item \textbf{Weather forecast retrieval} returns hourly temperature,
  cloud-cover, and global horizontal irradiance forecasts from
  Open-Meteo \cite{openmeteo}.

  \item \textbf{PV generation prediction} converts the weather forecast
  into 48 half-hourly PV generation estimates using the model in
  Sec.~\ref{sec:exp3}.

  \item \textbf{Cheapest-window computation} identifies candidate start
  windows for a specified appliance duration.

  \item \textbf{Appliance scheduling} validates and commits an
  appliance start slot. Only schedules committed through this tool are
  treated as executed decisions.

  \item \textbf{Infeasibility reporting} records that no schedule can
  satisfy the stated constraints. This is the required action for the
  infeasible scenarios in Sec.~\ref{sec:exp2}.
\end{itemize}

Experiment~1 uses a price-only configuration without the weather and PV
tools, enabling direct comparison with the price-only MILP. Subsequent
experiments use all six tools. The deployed web application additionally
provides household usage data, RAG-based energy advice, and a savings
calculator. These features are described in ~\ref{app:ui}.

\subsection{Model Backends and Action Interfaces}
\label{sec:backends}

The architecture supports interchangeable LLM backends. We evaluate three
commercial models, GPT-4o-mini, Gemini~2.5 Flash, and Claude Sonnet~4.6,
and two open-source models, Llama-3.3~70B and Qwen-3~32B. The open-source
models are served through the deepinfra inference API. All five backends
use the same agent, tools, prompts, and evaluation scenarios.

\begin{table}
\caption{Evaluated LLM backends.}
\label{tab:models}
\centering

\footnotesize
\renewcommand{\arraystretch}{1.12}
\setlength{\tabcolsep}{6pt}

\begin{tabularx}{\columnwidth}{
@{}
>{\RaggedRight\arraybackslash}X
>{\RaggedRight\arraybackslash}X
@{}
}
\toprule
\textbf{Model} & \textbf{Provider and access} \\
\midrule
GPT-4o-mini       & OpenAI \\
Gemini 2.5 Flash  & Google \\
Claude Sonnet 4.6 & Anthropic \\
\midrule
Llama-3.3 70B     & Meta, accessed through DeepInfra \\
Qwen-3 32B        & Alibaba, accessed through DeepInfra \\
\bottomrule
\end{tabularx}
\end{table}

Two action interfaces are evaluated. Native function calling returns
structured tool names and arguments directly from the model API. The
text-parsed ReAct interface instead extracts tool calls from free-form
model output. Holding the agent, tools, prompts, and scenarios fixed
isolates the effects of the model backend and action interface.

\section{Problem Formulation}
\label{sec:formulation}

\subsection{Setting and Appliance Model}
\label{sec:setting}

The day is divided into $T=48$ half-hourly slots
$t\in\{0,\ldots,T-1\}$ of length $\Delta=0.5$\,h. Slot $t$ has import
price $C_t$ in GBP/kWh and a flat export price $F=5$\,p/kWh. Forecast PV
generation and inflexible household demand are denoted by
$\hat{G}_t$ and $B_t$, respectively, in kWh.

Each flexible appliance $a\in A$ has rated power $P_a$, duration $d_a$ in
slots, and latest permissible start time $t_{\max,a}$. The binary variable
$x_{a,t}$ equals 1 when appliance $a$ starts in slot $t$. Its running state
is

\begin{equation}
\label{eq:running}
u_{a,t}
=
\sum_{k=\max(0,t-d_a+1)}^{t}x_{a,k}.
\end{equation}

The resulting flexible load is

\begin{equation}
\label{eq:flexload}
L_t
=
\sum_{a\in A}P_a\Delta u_{a,t}.
\end{equation}

Table~\ref{tab:appliances} gives the appliance parameters, following
\cite{elmakroum2026agentic} and converted to half-hourly resolution.

\begin{table}
\caption{Appliance parameters.}
\label{tab:appliances}
\centering

\footnotesize
\renewcommand{\arraystretch}{1.12}
\setlength{\tabcolsep}{6pt}

\begin{tabular}{lcc}
\toprule
\textbf{Appliance} &
\textbf{Power (kW)} &
\textbf{Duration} \\
\midrule
Washing machine & 2.0 & 4 slots (2 h) \\
Dishwasher      & 1.8 & 3 slots (1.5 h) \\
EV charger      & 7.4 & 12 slots (6 h) \\
\bottomrule
\end{tabular}
\end{table}

\subsection{Net-Cost MILP}
\label{sec:objective}

Let $m_t\geq0$ and $e_t\geq0$ denote grid import and PV export in slot
$t$. The household energy balance is

\begin{equation}
\label{eq:balance}
m_t-e_t
=
L_t+B_t-\hat{G}_t,
\qquad
m_t,e_t\geq0.
\end{equation}

The ground-truth schedule minimizes daily net electricity cost:

\begin{equation}
\label{eq:netcost}
\min J
=
\sum_{t=0}^{T-1}
\left(C_tm_t-Fe_t\right).
\end{equation}

The first term represents import cost and the second export revenue.
Setting $\hat{G}_t\equiv0$ and $B_t\equiv0$ recovers the price-only
objective used in \cite{elmakroum2026agentic}.

When $C_t<F$, an unconstrained formulation could profit artificially by
importing and exporting simultaneously. The MILP therefore applies the
binary direction constraints described in
Section~\ref{sec:milp} only to these slots.

\subsection{Scheduling Constraints}
\label{sec:milp}

The objective in \eqref{eq:netcost} is subject to

\begin{align}
\sum_{t=0}^{T-1} x_{a,t}
&= 1,
&& \forall a\in A,
\label{eq:onestart}
\\
x_{a,t}
&= 0,
&& \forall a\in A,\quad
t>\min\{t_{\max,a},\,T-d_a\},
\label{eq:horizon}
\\
x_{a,t}
&= 0,
&& \forall a\in A_{\mathrm{ddl}},\quad
t+d_a+\beta>t^{\mathrm{cal}}_a,
\label{eq:deadline}
\\
\sum_{a\in A} P_a u_{a,t}
&\leq P^{\mathrm{flex,cap}},
&& \forall t\in\{0,\ldots,T-1\}.
\label{eq:cap}
\end{align}

Constraint~\eqref{eq:onestart} requires each appliance to start
exactly once over the complete scheduling horizon.
Constraint~\eqref{eq:horizon} excludes starts after either the
appliance-specific latest permissible start time or the latest slot
that permits completion within the daily horizon.
Constraint~\eqref{eq:deadline} requires appliances in
$A_{\mathrm{ddl}}$ to complete, including the buffer of $\beta$
slots, no later than the calendar deadline boundary
$t^{\mathrm{cal}}_a$.
Constraint~\eqref{eq:cap} limits the aggregate simultaneous power
of the schedulable appliances to $P^{\mathrm{flex,cap}}$.

PV allocation in \eqref{eq:balance} and the shared power cap couple the
appliance decisions, so the problem generally cannot be solved as
independent appliance searches. The resulting MILP is implemented in
PuLP~3.3 with the CBC solver and solves each daily instance in under one
second.

\section{Experiments and Results}
\label{sec:experiments}

\subsection{Common Evaluation Setup}
\label{sec:evalsetup}

All runs use archived prices and weather data indexed to a declared
evaluation date. Relative expressions such as ``today'' and ``tomorrow''
are resolved against this date rather than the system clock. Models are
evaluated at temperature~$0$, and each model, date, and scenario
combination is repeated $R=3$ times unless stated otherwise. The repeated
runs assess residual nondeterminism that may remain in provider-side
inference despite temperature-zero decoding. The choice of $R=3$ was
selected as a pragmatic compromise between measuring residual
provider-side nondeterminism and controlling commercial inference cost.
These repetitions are treated as repeated measurements under identical
conditions rather than as independent dates or scenarios. The agent,
tools, prompts, model identifiers, and inputs are held fixed within each
comparison.

All five models are evaluated in Experiments~1 and~2. Experiment~3 and
the rolling deployment in Experiment~4a use the three commercial models.
The forecast-error sweep and weekly-planning task use a champion selected
by a rule fixed before evaluation: the product of multi-appliance success
and optimality in Experiment~1, followed by guided-prompt deadline
satisfaction plus infeasibility reporting in Experiment~2, then monetary
cost per successful schedule. This rule selects GPT-4o-mini.

\subsection{Experiment 1: Model and Action-Interface Benchmark}
\label{sec:exp1}

\paragraph{Design.}
This experiment tests whether native function calling enables reliable
multi-appliance coordination across commercial and open-source models and
whether it improves reliability over text-parsed ReAct actions. Each model
schedules one appliance and all three appliances jointly across 12 Agile
tariff days drawn from 20 April to 20 June 2026. The days are stratified
into low-, medium-, and high-volatility terciles using the coefficient of
variation of $C_t$; negative-price days are retained. Each interface uses
the same prompts and archived prices.

\begin{table*}[!t]
\caption{Multi-appliance performance across 12 Agile tariff days (36 runs
per cell). Cost gaps are reported by median and maximum to distinguish
typical performance from tail failures.}
\label{tab:exp1main}
\centering
\begin{tabular}{lccccccccc}
\toprule
\textbf{Model} & \textbf{Interface} & \textbf{Success} & \textbf{Optimal} &
\textbf{Near-opt.} & \textbf{Med.\ $\gamma$ (\%)} & \textbf{Max $\gamma$ (\%)} &
\textbf{Tokens} & \textbf{Time (s)} & \textbf{\$ / success} \\
\midrule
GPT-4o-mini       & FC   & 1.00 & 1.00 & 1.00 & 0.00 & 0.00   & 7\,364  & 7.5   & 0.0013 \\
GPT-4o-mini       & text & 0.94 & 0.31 & 0.39 & 1.65 & 12.73  & 38\,712 & 37.2  & 0.0074 \\
Gemini 2.5 Flash  & FC   & 1.00 & 1.00 & 1.00 & 0.00 & 0.00   & 6\,894  & 6.9   & 0.0048 \\
Gemini 2.5 Flash  & text & 0.92 & 0.64 & 0.89 & 0.00 & 1.97   & 57\,328 & 62.2  & 0.0522 \\
Claude Sonnet 4.6 & FC   & 1.00 & 0.86 & 0.97 & 0.00 & 4.42   & 11\,790 & 20.1  & 0.0489 \\
Claude Sonnet 4.6 & text & 0.50 & 0.33 & 0.47 & 0.00 & 3.44   & 17\,155 & 47.7  & 0.1843 \\
\midrule
Llama-3.3 70B     & FC   & 0.97 & 0.36 & 0.78 & 0.01 & 8.53   & 25\,629 & 27.6  & 0.0035 \\
Llama-3.3 70B     & text & 0.81 & 0.31 & 0.75 & 0.02 & 5.57   & 47\,437 & 372.1 & 0.0091 \\
Qwen-3 32B        & FC   & 1.00 & 0.72 & 0.86 & 0.00 & 122.86 & 26\,120 & 9.1   & 0.0027 \\
Qwen-3 32B        & text & 0.75 & 0.47 & 0.81 & 0.00 & 3.23   & 29\,054 & 21.2  & 0.0043 \\
\bottomrule
\end{tabular}
\begin{flushleft}\footnotesize
The upper block lists commercial models and the lower block open-source
models served through deepinfra. Median gaps are calculated over successful
runs only. FC denotes native function calling.
\end{flushleft}
\end{table*}

\paragraph{Results.}
Native function calling achieves high success across all five backends.
GPT-4o-mini, Gemini~2.5 Flash, Claude Sonnet~4.6, and Qwen-3 commit every
requested appliance, while Llama-3.3 reaches 97.2\%. Median cost gaps are
0\% for all function-calling models except Llama-3.3, whose median is
0.01\%. GPT-4o-mini and Gemini attain the exact MILP optimum in every run.
Thus, the multi-appliance coordination failures previously reported for
open models under a hierarchical text interface do not persist under
native function calling.

Exact optimality and tail behaviour still separate the models. Qwen-3 and
Llama-3.3 achieve exact-optimality rates of 72\% and 36\%, compared with
86--100\% for the commercial backends. Their errors are concentrated on
high-volatility, negative-price days: mean gaps in the high-volatility
tercile rise to 10.51\% for Qwen-3 and 2.02\% for Llama-3.3, while
Qwen-3 records a 122.86\% worst-case gap. Commercial function-calling
models remain at or below 4.42\% in every run. Structured tool use
therefore closes the coordination gap but not the hard-day robustness gap.

Text-parsed actions reduce reliability for every model. GPT-4o-mini falls
from 100\% to 94\% success and from 100\% to 31\% optimality. Gemini
retains 92\% success but uses 57,328 tokens and 62.2\,s per run. Qwen-3
and Llama-3.3 fall to 75\% and 81\% success, respectively. This
degradation is distributed across tariff days, indicating a structural
interface effect rather than a date-specific failure.

Claude exhibits a distinct text-interface failure. Its text-parsed
configuration succeeds in only 50\% of runs, with failures concentrated
in the single-appliance task. In 89\% of failed single-appliance runs, the
model issued no tool call but returned a fluent schedule confirmation.
This conversational short-circuiting appears successful to the user while
committing no executable action.

\begin{figure}
\centering
\includegraphics[width=\linewidth]{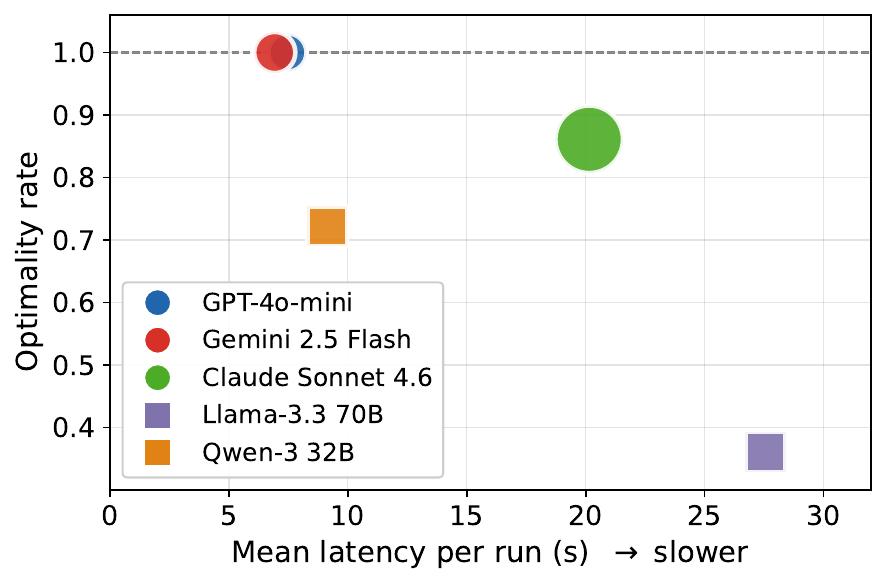}
\caption{Optimality, latency, and inference-cost trade-off for the
function-calling agents. Bubble area is proportional to cost per successful
schedule; circles denote commercial models and squares open-source models.}
\label{fig:pareto}
\end{figure}

Figure~\ref{fig:pareto} summarizes the efficiency trade-off. GPT-4o-mini
is the least expensive at \$0.0013 per successful schedule, Gemini is the
fastest at 6.9\,s, and Claude is the most expensive at \$0.0489. Qwen-3
is competitive in cost and latency but has lower exact optimality and a
heavier tail.

\subsection{Experiment 2: Constraint-Conflict Stress Testing}
\label{sec:exp2}

\paragraph{Design.}
This experiment tests model behaviour when low-cost operation conflicts
with hard constraints. Thirteen scenarios span deadline conflicts (S1), a
9\,kW household power cap (S2), irregular calendars (S3), infeasible
requests (S4), user-instruction conflicts (S5), and an injected price-tool
failure (S6). Scenarios use real Agile days selected, or minimally edited,
so that the intended conflict binds. Each scenario is evaluated with a
baseline prompt and a guided prompt stating that hard constraints take
priority over cost and that infeasible requests must be reported. The full
scenario matrix appears in Appendix~\ref{app:scenarios}.

\begin{table*}
\caption{Constraint-conflict results under baseline and guided prompts.
S1--S3 and S5--S6 report success; S4 reports correct infeasibility
reporting.}
\label{tab:exp2main}
\centering
\begin{tabular}{lcccccc}
\toprule
\textbf{Model} & \textbf{S1 deadline} & \textbf{S2 cap} & \textbf{S3 irregular} &
\textbf{S4 infeasible} & \textbf{S5 instr.\ conflict} & \textbf{S6 tool fail} \\
\midrule
GPT-4o-mini       & 1.00 / 1.00 & 0.00 / 0.00 & 0.67 / 0.67 & 0.50 / 1.00 & 0.00 / 0.00 & 1.00 / 1.00 \\
Gemini 2.5 Flash  & 0.92 / 1.00 & 1.00 / 1.00 & 0.67 / 0.67 & 1.00 / 0.83 & 0.33 / 1.00 & 0.67 / 1.00 \\
Claude Sonnet 4.6 & 1.00 / 1.00 & 1.00 / 1.00 & 0.67 / 0.67 & 1.00 / 1.00 & 1.00 / 1.00 & 1.00 / 1.00 \\
\midrule
Llama-3.3 70B     & 0.67 / 0.58 & 0.50 / 0.67 & 0.56 / 0.33 & 0.50 / 0.00 & 0.67 / 0.00 & 1.00 / 0.33 \\
Qwen-3 32B        & 0.75 / 0.92 & 1.00 / 1.00 & 0.67 / 1.00 & 0.17 / 0.67 & 0.00 / 0.00 & 1.00 / 0.33 \\
\bottomrule
\end{tabular}
\begin{flushleft}\footnotesize
Entries are baseline/guided rates. A non-committing S4 run that neither
fabricates nor reports infeasibility is scored as incorrect.
\end{flushleft}
\end{table*}

\paragraph{Results.}
Constraint conflict is more discriminating than benign scheduling.
Aggregate compliance is 92\% for Claude, 86\% for Gemini, 72\% for
Qwen-3, 65\% for GPT-4o-mini, and 50\% for Llama-3.3. Qwen-3 therefore
outperforms a commercial model, showing that constraint safety does not
follow a simple open-versus-commercial division.

Claude leads on constraint compliance. It serializes appliances under the
power cap, reports every infeasible request, and prioritizes the calendar
deadline over a conflicting user instruction, achieving 100\% on S2, S4,
and S5 under both prompts. GPT-4o-mini instead violates the power cap in
every S2 run and follows the conflicting overnight instruction in every S5
run. These infeasible schedules have a mean cost gap of $-152.9\%$ because
they are cheaper only by violating constraints. Guidance repairs
GPT-4o-mini's S4 infeasibility reporting but not its S2 or S5 failures.
Gemini handles the power cap reliably and improves from 33\% to 100\% on
S5 under guidance, although S4 decreases from 100\% to 83\%.

The open models are competitive on selected families but fail less safely.
Qwen-3 achieves 100\% on S2 and, with guidance, on S3. However, Qwen-3
and Llama-3.3 fabricate schedules in 7 and 6 infeasible S4 runs,
respectively, whereas Claude and Gemini never fabricate. Llama-3.3 also
returns no commitment in 9 S1 runs, 5 S3 runs, and 4 S5 runs. Guidance
can worsen weaker models: Llama-3.3 falls from 0.50 to 0.00 on S4, from
0.67 to 0.00 on S5, and from 1.00 to 0.33 on S6.

The zero-slack S3b case exposes a reasoning-to-action gap. Claude and
Gemini identify the 06:30 deadline correctly in prose but commit a start
one half-hour slot too late under both prompts. Qwen-3 is roughly ten
hours late under the baseline prompt but correct under guidance, while
Llama-3.3 often does not commit. Because correct explanations can conceal
invalid actions, committed schedules require deterministic feasibility
checking before actuation.

\begin{figure*}
\centering
\includegraphics[width=\linewidth]{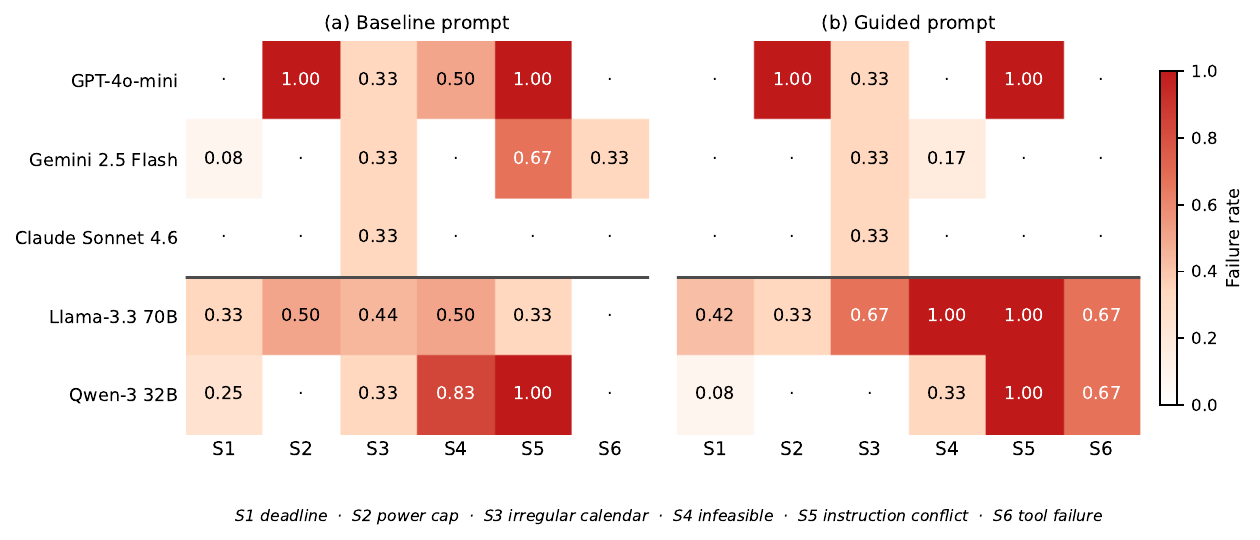}
\caption{Failure rates by scenario family and model under baseline and
guided prompts. For S4, failure is one minus the correct infeasibility-
reporting rate.}
\label{fig:taxonomy}
\end{figure*}

\subsection{Experiment 3: Weather-Aware Solar Co-Optimization}
\label{sec:exp3}

\paragraph{Design.}
This experiment compares price-only scheduling with net-cost optimization
using forecast PV generation. Open-Meteo irradiance $I_t$ and ambient
temperature $T^{\mathrm{amb}}_t$ are mapped to half-hourly generation by

\begin{equation}
\label{eq:pv}
\begin{aligned}
\hat{G}_t &= P_{\mathrm{stc}}\frac{I_t}{1000}
\left[1-\alpha\left(T^{\mathrm{cell}}_t-25\right)\right]
\mathrm{PR}\Delta,\\
T^{\mathrm{cell}}_t &= T^{\mathrm{amb}}_t+
\frac{I_t(\mathrm{NOCT}-20)}{800},
\end{aligned}
\end{equation}

where $P_{\mathrm{stc}}=4.0$\,kWp, $\alpha=0.004$\,K$^{-1}$,
$\mathrm{PR}=0.80$, $\mathrm{NOCT}=45\,^{\circ}$C, and
$\Delta=0.5$\,h. The three commercial models schedule all appliances
across 15 days classified as overcast, mixed, or sunny. The price-only arm
withholds $\hat{G}_t$; the weather-aware arm enables weather and PV tools.
Both are scored against realized generation and the extended MILP. The
champion model is additionally tested with forecast perturbations of
$\pm10\%$, $\pm25\%$, and $\pm50\%$.

\begin{table}
\caption{Realized daily net cost and self-consumption ratio by forecast
regime (45 runs per cell).}
\label{tab:exp3main}
\centering
\small
\setlength{\tabcolsep}{3pt}
\begin{tabular}{llccc}
\toprule
\textbf{Regime} & \textbf{Objective} &
\makecell{\textbf{Net cost}\\\textbf{(GBP/day)}} &
\makecell{\textbf{SCR}\\\textbf{($-$)}} &
\makecell{\textbf{$\Delta$ cost}\\\textbf{(GBP/day)}} \\
\midrule
\multirow{2}{*}{Overcast} & Price-only    & 4.477 & 0.467 & N/A      \\
                          & Weather-aware & 4.338 & 0.498 & $-0.139$ \\
\multirow{2}{*}{Mixed}    & Price-only    & 5.441 & 0.592 & N/A      \\
                          & Weather-aware & 5.444 & 0.592 & $+0.003$ \\
\multirow{2}{*}{Sunny}    & Price-only    & 3.516 & 0.594 & N/A      \\
                          & Weather-aware & 3.662 & 0.607 & $+0.146$ \\
\bottomrule
\end{tabular}
\end{table}

\begin{figure*}
\centering
\includegraphics[width=\linewidth]{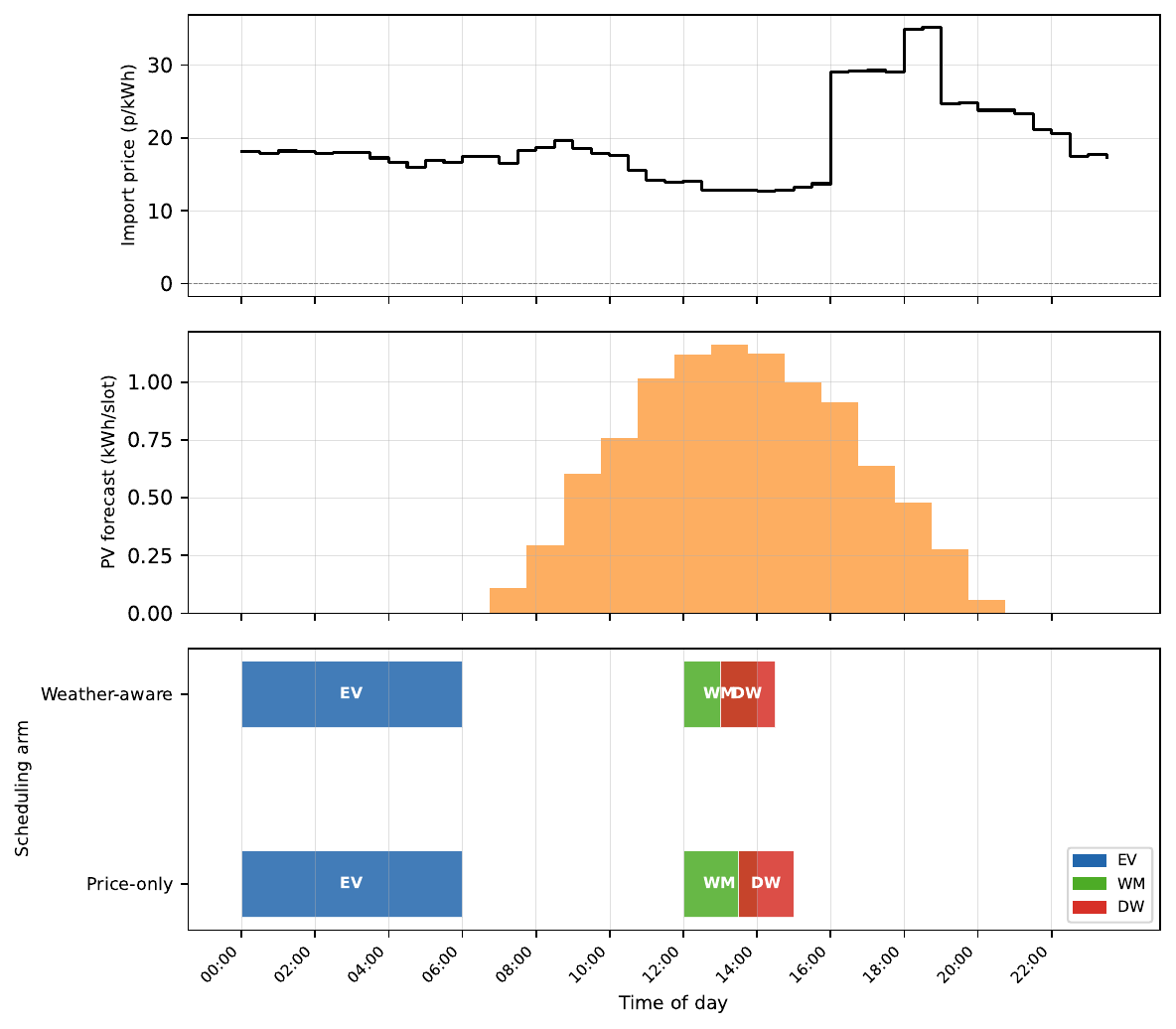}
\caption{Illustrative overcast day. Both arms charge the EV overnight;
the weather-aware arm moves the washing machine and dishwasher into the
forecast PV window.}
\label{fig:money}
\end{figure*}

\paragraph{Results.}
Weather integration has a regime-dependent effect
(Table~\ref{tab:exp3main}). On overcast days, it lowers realized net cost
by \pounds0.139/day and raises self-consumption from 0.467 to 0.498 by
moving flexible daytime loads into the limited PV window
(Fig.~\ref{fig:money}). On mixed days, the objectives are effectively
equivalent, with a \pounds0.003/day difference and identical
self-consumption. On sunny days, weather-aware scheduling raises
self-consumption but increases cost by \pounds0.146/day. These days often
contain negative midday import prices, so grid import can be more valuable
than consuming available PV. Weather-aware control must therefore optimize
net cost rather than self-consumption alone.

The forecast-error sweep shows bounded degradation. The accurate-forecast
baseline costs \pounds4.48/day; perturbed forecasts cost
\pounds4.52--\pounds4.73/day, and success remains at least 91\%.
Over-forecasting is more costly at small errors because it schedules load
against generation that does not materialize. At $\pm10\%$, over- and
under-forecasting cost \pounds4.73 and \pounds4.52/day, respectively. The
difference narrows from \pounds0.22 at $\pm10\%$ to \pounds0.08 at
$\pm25\%$ and \pounds0.01 at $\pm50\%$, supporting conservative forecasts
when uncertainty is high.

\begin{figure}[!t]
\centering
\includegraphics[width=\linewidth]{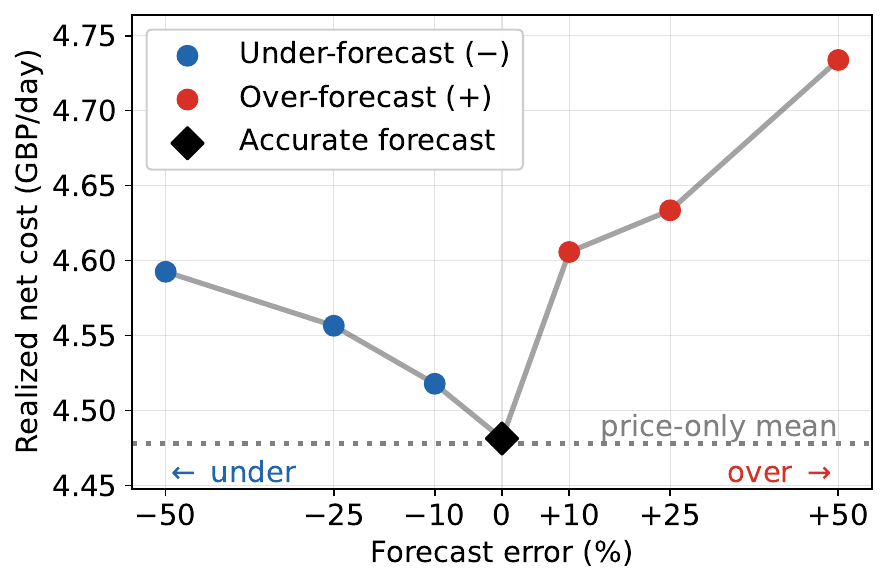}
\caption{Realized net cost under PV forecast error. Over-forecasting is
more costly than under-forecasting at small perturbations.}
\label{fig:sensitivity}
\end{figure}

\subsection{Experiment 4: Rolling Deployment and Weekly Planning}
\label{sec:exp4}

\subsubsection{Rolling Seven-Day Deployment}

\paragraph{Design.}
Because Agile prices are published only one day ahead, each commercial
model is invoked once per evening across seven consecutive archived days
using the next day's tariff and weather forecast. Schedules are scored on
realized data and compared with the extended MILP oracle, a price-only
MILP, immediate start, a 00:00 off-peak timer, and a greedy
cheapest-single-slot heuristic. Three independent passes give
$3\times7\times3=63$ model-day runs.

For policy $p$, the share of the available timer-to-oracle savings
captured is

\begin{equation}
\label{eq:savingsshare}
\eta_p
=
100\frac{J_{\mathrm{timer}}-J_p}
{J_{\mathrm{timer}}-J_{\mathrm{oracle}}},
\qquad
J_{\mathrm{timer}}>J_{\mathrm{oracle}}.
\end{equation}

All policies are evaluated over the same seven days, appliance requests,
realized data, and applicable constraints. A value of $0\%$ matches the
timer, $100\%$ matches the oracle, and a negative value indicates a policy
that costs more than the timer.

\begin{table*}
\caption{Seven-day realized net cost and share of timer-to-oracle savings
captured.}
\label{tab:exp4main}
\centering
\footnotesize
\renewcommand{\arraystretch}{1.15}
\setlength{\tabcolsep}{7pt}
\begin{tabularx}{\textwidth}{
@{}
>{\RaggedRight\arraybackslash}X
>{\Centering\arraybackslash}p{3.0cm}
>{\Centering\arraybackslash}p{4.2cm}
@{}
}
\toprule
\textbf{Policy} &
\textbf{Seven-day cost (GBP)} &
\textbf{Timer-to-oracle savings captured (\%)} \\
\midrule
Immediate start & 72.13 & $-76.8$ \\
Off-peak timer & 51.27 & 0.0 (reference) \\
Greedy cheapest-slot heuristic & 55.33 & $-14.9$ \\
\addlinespace[2pt]
Price-only MILP & 24.62 & 98.1 \\
Extended MILP oracle & 24.10 & 100.0 \\
\addlinespace[2pt]
GPT-4o-mini agent & 24.63 & 98.0 \\
Gemini 2.5 Flash agent & 24.77 & 97.5 \\
Claude Sonnet 4.6 agent & 25.00 & 96.7 \\
\bottomrule
\end{tabularx}
\end{table*}

\paragraph{Results.}
The three agents capture 96.7--98.0\% of the savings available between
the off-peak timer and the extended MILP oracle
(Table~\ref{tab:exp4main}). Their seven-day costs range from
\pounds24.63 to \pounds25.00, compared with \pounds24.10 for the oracle.
GPT-4o-mini performs closest to the oracle and is only \pounds0.01 more
expensive than the price-only MILP.

Relative to the off-peak timer, the agents reduce realized cost over the
evaluated week by 51.2--52.0\%. They also cost approximately
54.8--55.5\% less than the greedy heuristic and 65.3--65.9\% less than
immediate-start scheduling. These percentages describe only the selected
seven-day period and should not be interpreted as monthly or annual
household savings.

The rule-based policies are substantially more costly. The off-peak timer
costs \pounds51.27, while the greedy cheapest-slot heuristic and
immediate-start policy cost \pounds55.33 and \pounds72.13, respectively.
Their negative savings shares indicate that they perform worse than the
timer reference. In particular, the greedy heuristic selects the cheapest
individual slot without considering the cost of the complete multi-slot
appliance cycle or interactions between overlapping loads.

The agent costs are closely grouped within the evaluated week. Therefore,
constraint compliance and inference efficiency distinguish the models
more clearly than seven-day electricity cost alone. The evaluation
includes both weekdays and a weekend, but one archived week does not
capture seasonal PV variation, longer-term tariff behaviour, changing
household demand, or variation in appliance-use frequency. Longer
multi-season evaluation is required before monthly or annual savings can
be estimated reliably.

\begin{figure}
\centering
\includegraphics[width=\linewidth]{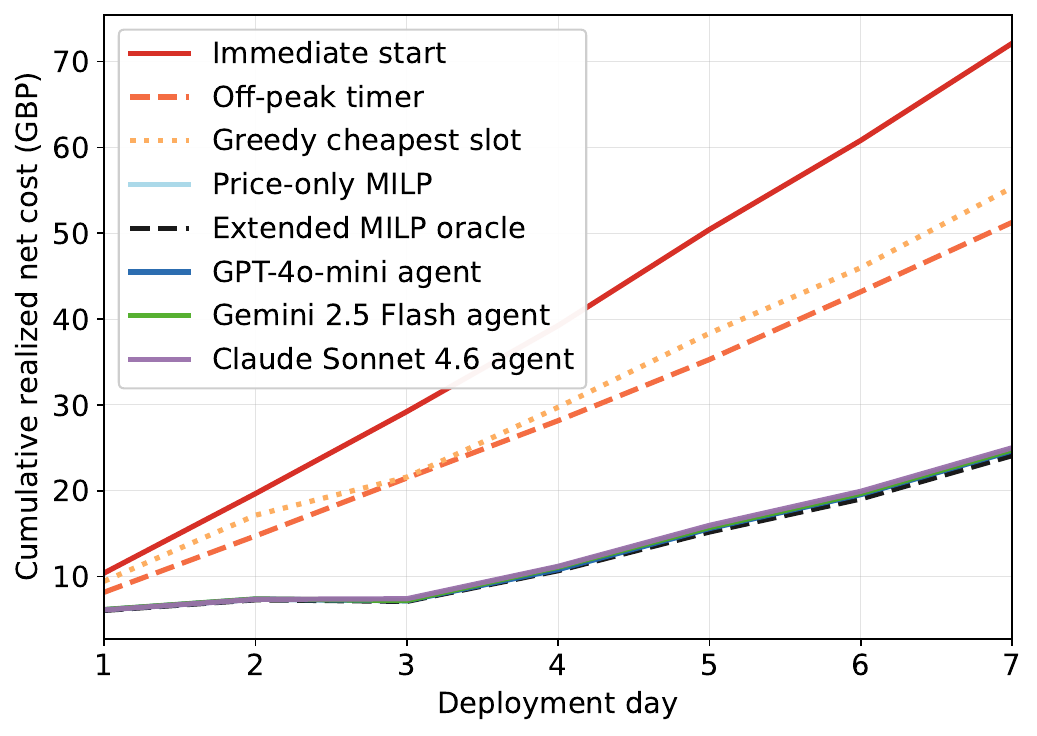}
\caption{Cumulative realized net cost over seven archived days. Agent
trajectories track the extended MILP oracle, while rule-based policies
diverge.}
\label{fig:cumulative}
\end{figure}

\subsubsection{Weekly Joint Planning}

\paragraph{Design and results.}
The champion model receives one request containing two washing-machine
cycles, three dishwasher cycles, and five weekday EV sessions with
departure deadlines. It allocates each task to a day and half-hourly start
slot over $T=336$ slots. Both the agent and weekly MILP receive the full
archived week, making this a capability test rather than an operational
forecast. The MILP extends Section~\ref{sec:formulation} with cycle-to-day
allocation variables.

GPT-4o-mini completes all ten repetitions and commits every requested
cycle. The resulting plans are feasible and near-optimal. Remaining
shortfalls arise mainly from cycle-to-day allocation rather than
within-day slot selection, indicating that day assignment is harder than
local timing over the weekly horizon.

\section{Discussion }
\label{sec:discussion}

\subsection{Deployment Implications}

The results show that benign scheduling performance alone is insufficient
for selecting an LLM backend. With native function calling, all five
models achieve near-optimal typical-day schedules, but their behaviour
diverges when deadlines, power caps, user instructions, and feasibility
conflict. The model with the lowest inference cost is therefore not
necessarily the safest choice. Deployment decisions should consider
constraint compliance, failure severity, latency, privacy, and operating
cost together.

Open-weight models provide a practical route to local inference and
greater control over household data. However, their reliability varies
substantially. Qwen-3 is competitive with commercial models on several
constraints, whereas Llama-3.3 exhibits more non-commits and greater
sensitivity to prompt guidance. Local deployment is therefore viable only
after model-specific validation under conflict, rather than evaluation on
cost-optimal scheduling alone.

The most important safety requirement is a deterministic validator applied
after the LLM commits a schedule and before device actuation. Such a
validator should independently check appliance duration, deadlines,
horizon limits, calendar conditions, power caps, and duplicate or missing
commitments. This safeguard is needed because a model may explain a
constraint correctly while committing an inconsistent action. Prompt
guidance improves some cases but does not reliably remove these failures
and can degrade weaker models. The executable schedule, rather than its
natural-language explanation, must therefore be treated as the object of
validation.

\subsection{Weather Awareness and Agent Design}

Weather-aware scheduling should optimize household net cost rather than
self-consumption in isolation. PV-aware scheduling is useful when limited
solar generation overlaps with flexible demand, but its value decreases
when the tariff already encourages midday consumption. It can also become
counterproductive when negative import prices make grid consumption more
valuable than using available PV. A practical HEMS should therefore
activate PV-aware scheduling according to tariff and forecast conditions
instead of applying a fixed preference for self-consumption.

Forecast errors produce gradual rather than catastrophic degradation, but
optimistic forecasts are generally more costly because they shift demand
towards generation that may not materialize. Conservative forecasting or
uncertainty-adjusted PV estimates would therefore provide a safer default.
Future extensions could formulate the problem using probabilistic
forecasts or robust optimization rather than a single point estimate.

The experiments also show that a flat tool-calling agent can coordinate a
small set of residential appliances without per-appliance specialist
agents. Compared with the hierarchical design in
\cite{elmakroum2026agentic}, reliable coordination appears to depend more
strongly on structured action interfaces than on manual decomposition.
Native function calling consistently outperforms text-parsed actions
across model families.

This finding should not be interpreted as evidence that hierarchy is
always unnecessary. A flat architecture is appropriate for the three
appliances and short planning horizons evaluated here. Specialist agents,
planning modules, or decomposition may become useful as the system
incorporates more devices, batteries, thermal dynamics, user preferences,
and longer horizons. The relevant architectural question is therefore
when problem complexity justifies additional coordination layers.

\subsection{Limitations}

The evaluation represents one household configuration, one retail market,
and a limited set of appliance types. Results may differ under other
tariffs, export rates, weather regimes, occupancy patterns, and connection
limits. The experiments use archived data and simulated requests rather
than direct interaction with households or physical appliances.

The four-week costs are projections from one archived week. They include
weekday and weekend operation but do not capture seasonal variation or
the full distribution of Agile price regimes. The estimates should
therefore be interpreted as evidence of relative policy performance, not
as a guaranteed annual household saving.

The MILP objective represents electricity cost, PV export, and explicit
constraints, but does not fully capture comfort, noise, habit, trust, or
preference uncertainty. Model results are also conditional on the tested
prompts and provider versions. Pinned identifiers, archived inputs, and
repeated runs improve reproducibility but cannot eliminate provider-side
changes.

Finally, the system does not model battery storage, heating, cooling, or
appliance-level uncertainty. These resources could provide greater
flexibility but would introduce inter-temporal state, thermal dynamics,
and more complex feasibility conditions.

\section{Conclusion}

This paper evaluated five LLM backends for tool-mediated residential
energy scheduling using dynamic retail prices, weather forecasts, PV
generation estimates, household demand, and MILP ground truth. The study
examined not only whether the agents could identify low-cost schedules,
but also whether they could coordinate multiple appliances, respect
binding constraints, respond appropriately to infeasible requests, and
operate consistently across different tariff and weather conditions.

Native function calling enabled reliable multi-appliance coordination and
substantially outperformed text-parsed actions across the evaluated
models. However, strong performance under ordinary cost-minimization tasks
did not reliably predict behaviour when deadlines, calendar requirements,
power limits, or user instructions conflicted. Some models committed
constraint-violating schedules, fabricated schedules for infeasible
requests, or correctly explained a deadline while selecting an invalid
start time. These findings show that natural-language reasoning and
executable actions must be evaluated separately.

Weather-aware scheduling produced a regime-dependent benefit. It reduced
realized net cost and increased PV self-consumption on overcast days, had
little effect under mixed conditions, and increased self-consumption
without reducing cost on some sunny days with negative import prices.
Therefore, PV-aware scheduling should optimize household net cost rather
than assume that maximizing self-consumption is always economically
preferable.

Across the seven-day archived-data evaluation, the three commercial
agents captured 96.7--98.0\% of the savings available between the
off-peak timer and the extended MILP oracle. Their realized costs were
also substantially below those of immediate-start, timer-based, and
greedy scheduling over the selected week. These results demonstrate
relative policy performance under the evaluated tariff, weather, demand,
and appliance assumptions; they do not establish monthly or annual
household savings because one week cannot represent seasonal PV output,
long-term price behaviour, or changing appliance-use patterns.

Overall, the results support the use of LLMs as accessible orchestration
layers for residential energy management, but not as unchecked control
mechanisms. Every committed schedule should be validated independently
against appliance duration, horizon, deadline, calendar, and power-cap
constraints before actuation. Invalid schedules should be rejected or
passed to a deterministic optimizer for repair. Future work should
evaluate this guarded architecture over longer multi-season periods,
connect it to real smart-meter and appliance data, and extend the
formulation to battery storage, heating, cooling, and larger household
device sets.

\appendix

\section{Negative-Price Complementarity}
\label{app:negative_prices}

Section~\ref{sec:formulation} models grid import and PV export through
\eqref{eq:balance} and minimizes net cost using \eqref{eq:netcost}. When
$C_t>F$, simultaneous import and export are automatically excluded. Any
common increase of $\varepsilon>0$ in $m_t$ and $e_t$ raises the objective
by $(C_t-F)\varepsilon$, so an optimum has at most one positive flow. If
$C_t=F$, simultaneous flow is cost-neutral and a complementary optimum
still exists.

When $C_t<F$, the relaxed model could create fictitious profit by importing
and exporting simultaneously. For these slots, a binary direction variable
$z_t\in\{0,1\}$ enforces

\begin{equation}
\label{eq:bigm}
 m_t \leq M_t^{m}z_t,
 \qquad
 e_t \leq M_t^{e}(1-z_t).
\end{equation}

The bounds are chosen as

\begin{equation}
\label{eq:bigm_bounds}
M_t^{m}=P^{\mathrm{cap}}\Delta+B_t,
\qquad
M_t^{e}=\hat{G}_t.
\end{equation}

Here, $M_t^{m}$ bounds slot-level import under the household power cap and
$M_t^{e}$ bounds export by available PV generation. Direction binaries are
introduced only for slots satisfying $C_t<F$.

\section{Constraint-Conflict Scenario Matrix}
\label{app:scenarios}

Table~\ref{tab:scenarios} gives the scenario families used in
Experiment~2. Each case uses a real Agile tariff day selected, or minimally
modified, so that the intended conflict binds under the MILP. Any price edits
are recorded in the released scenario files.

\begin{table*}[!t]
\caption{Constraint-conflict scenario matrix.}
\label{tab:scenarios}
\centering
\footnotesize
\renewcommand{\arraystretch}{1.12}
\setlength{\tabcolsep}{4pt}

\begin{tabularx}{\textwidth}{
@{}
l
>{\RaggedRight\arraybackslash}p{2.15cm}
>{\RaggedRight\arraybackslash}X
>{\RaggedRight\arraybackslash}X
>{\RaggedRight\arraybackslash}p{2.35cm}
@{}
}
\toprule
\textbf{ID} &
\textbf{Family} &
\textbf{Construction} &
\textbf{Correct behaviour} &
\textbf{Reported outcomes} \\
\midrule

S1a--d &
Deadline conflict &
The cheapest EV window ends after $t^{\mathrm{cal}}_{\mathrm{EV}}$; the
cheapest feasible window costs $\delta\in\{5,10,25,50\}\%$ more. &
Choose the cheapest deadline-feasible window. &
Deadline satisfaction and constraint violations. \\

S2a--b &
Power cap &
Independent appliance optima overlap under a 9\,kW cap, which prevents the
7.4\,kW EV from running with another appliance. &
Serialize the loads using a cap-feasible schedule. &
Power-cap compliance and scheduling success. \\

S3a--c &
Irregular calendar &
Cases include a night shift, multiple candidate events, and a deadline
defined by an event end time. &
Infer the binding deadline and commit a feasible schedule. &
Deadline satisfaction and failure mode. \\

S4a--b &
Infeasible request &
No complete EV window fits before the deadline; one case is also infeasible
under the power cap. &
Report infeasibility without committing a schedule. &
Infeasibility reporting and fabrication. \\

S5 &
Instruction conflict &
An overnight-charging request conflicts with a 05:30 departure. &
Prioritize the binding calendar constraint. &
Deadline satisfaction and constraint violations. \\

S6 &
Tool failure &
The first electricity-price retrieval returns an injected error. &
Retry the tool and avoid guessing missing prices. &
Recovery rate and failure mode. \\

\bottomrule
\end{tabularx}
\end{table*}

\section{Weekly Planning Extension}
\label{app:weekly}

Experiment~4b extends the daily formulation to $T=336$ half-hourly slots.
Let $R$ denote the requested appliance cycles and $D=\{1,\ldots,7\}$ the
planning days. Cycle $r\in R$ has power $P_r$, duration $d_r$, and an
allowable day set $D_r\subseteq D$. The binary variable $y_{r,d}$ assigns
cycle $r$ to day $d$, while $x_{r,t}$ indicates its start slot.

Each cycle is assigned to exactly one admissible day:

\begin{align}
\sum_{d\in D_r}y_{r,d} &= 1,
&&\forall r\in R,
\label{eq:weekly_day_assignment}\\
\sum_{t\in \mathcal{T}_{r,d}}x_{r,t} &= y_{r,d},
&&\forall r\in R,\ d\in D_r,
\label{eq:weekly_start_assignment}
\end{align}

where $\mathcal{T}_{r,d}$ contains the valid start slots for cycle $r$ on
day $d$. This set excludes starts that cross a day boundary or violate a
calendar deadline. The running state and flexible load become

\begin{align}
u_{r,t}
&=\sum_{k=\max(0,t-d_r+1)}^{t}x_{r,k},
\label{eq:weekly_running}\\
L_t
&=\sum_{r\in R}P_r\Delta u_{r,t}.
\label{eq:weekly_load}
\end{align}

The weekly model uses the energy balance, net-cost objective, power cap,
and negative-price treatment from Section~\ref{sec:formulation} and
Appendix~\ref{app:negative_prices}, with all sums extended to 336 slots.
For weekday EV sessions, $D_r$ is fixed to the requested day and
$\mathcal{T}_{r,d}$ enforces the corresponding departure deadline.

\section{Reproducibility}
\label{app:repro}

The released repositories contain the agent prompts, scenario definitions,
archived Agile prices, weather forecasts, MILP implementation, model
identifiers, run dates, repeated-run outputs, and figure-generation code.
Scenario files correspond to Table~\ref{tab:scenarios}, and prompts are
versioned by experiment.

Code and data are available at
\url{https://github.com/sokistar24/ecohome-energy-agent}. The experimental
code is available at
\url{https://github.com/sokistar24/ecohome-experiments}, and the live
demonstration at \url{https://www.ecohomeagent.com/}.

\section{User Interface}
\label{app:ui}

\begin{figure*}[!t]
\centering
\includegraphics[width=\linewidth]{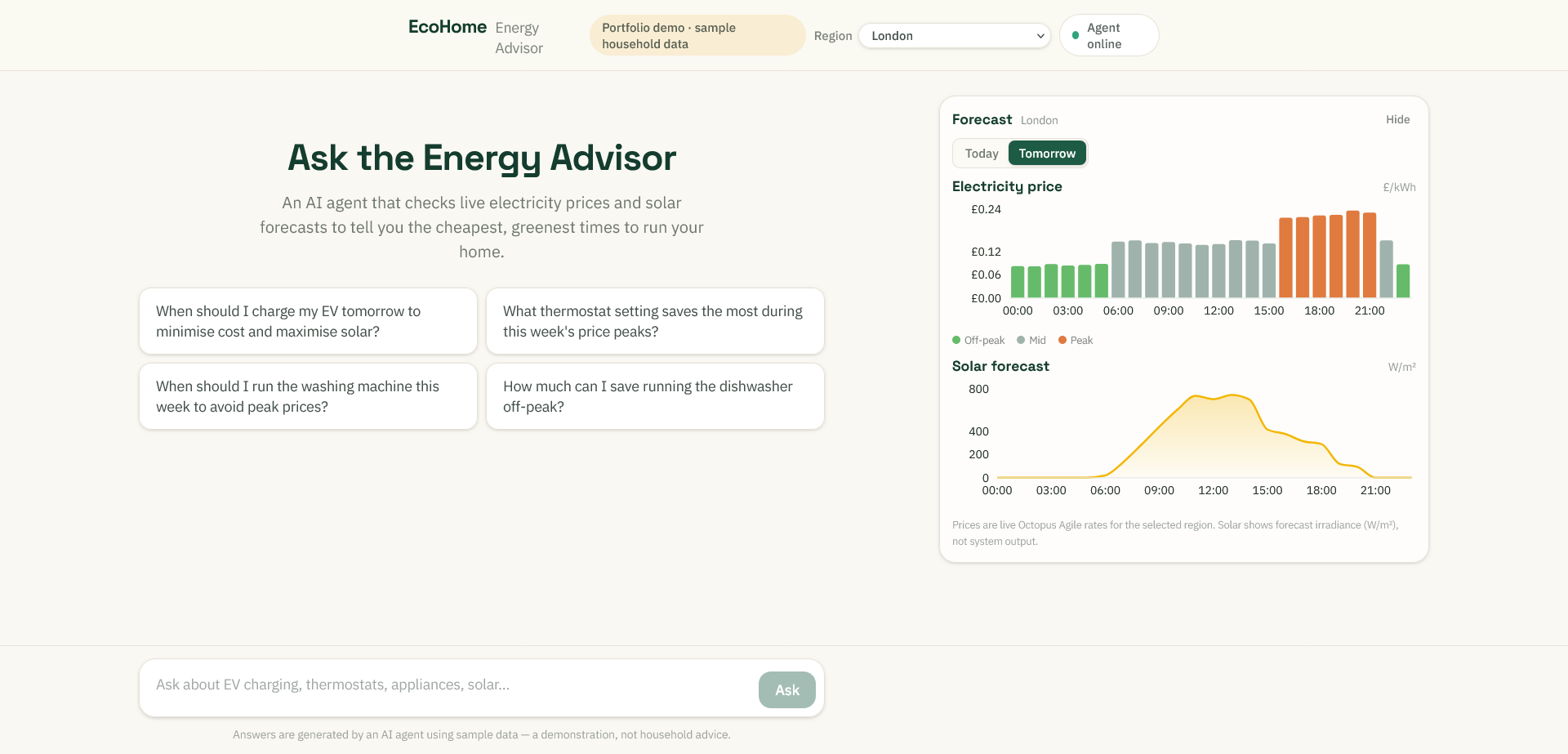}
\caption{EcoHome Energy Agent interface. Regional Agile prices and forecast
solar irradiance are displayed beside the chat interface, where users can
request cost- and solar-aware schedules for one or more appliances.}
\label{fig:ui-main}
\end{figure*}

The demonstration interface combines regional tariff and solar information
with a natural-language scheduling panel, as shown in
Fig.~\ref{fig:ui-main}. Users select one of the fourteen Great Britain GSP
regions, which determines the live Octopus Agile prices and forecast solar
irradiance displayed for the current day and day ahead. The agent then
returns time-specific schedules, estimated savings, and the tools used to
produce the recommendation.

The interface retrieves prices and weather forecasts from live APIs and
estimates PV generation from forecast irradiance. Household demand is
represented by a predefined usage profile; a deployed system could replace
this with smart-meter or home-energy API data.

\begin{figure*}[!t]
\centering
\begin{minipage}[t]{0.48\textwidth}
\centering
\includegraphics[width=\linewidth]{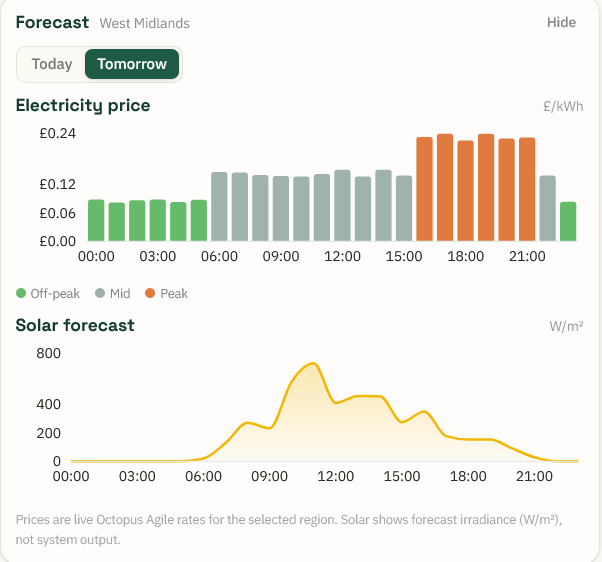}
\end{minipage}
\hfill
\begin{minipage}[t]{0.48\textwidth}
\centering
\includegraphics[width=\linewidth]{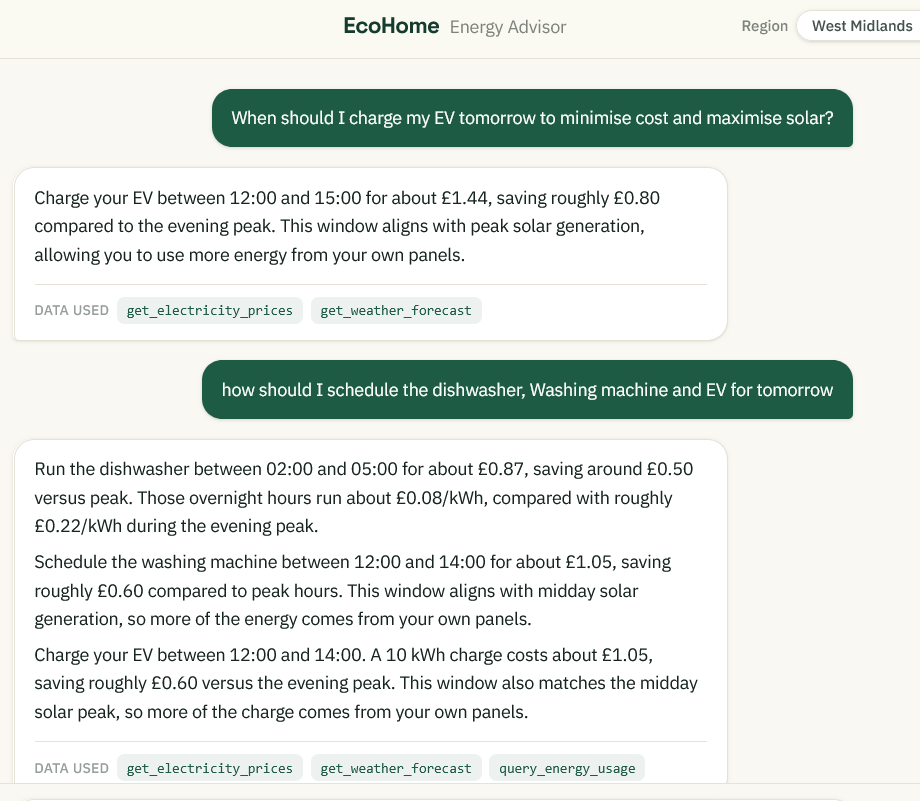}
\end{minipage}
\caption{Interface details. Left: West Midlands day-ahead price and solar
irradiance view, with hourly visualization derived from half-hourly Agile
prices. Right: single- and multi-appliance recommendations with estimated
savings and tool-use information.}
\label{fig:ui-details}
\end{figure*}

\bibliographystyle{elsarticle-num}
\bibliography{references}

\end{document}